\documentclass[a4paper,10pt]{article}
\usepackage[utf8]{inputenc}
\usepackage{graphicx}
\usepackage{natbib}
\usepackage{aas_macros}
\setcitestyle{authoryear,open={(},close={)}}
\bibpunct{(}{)}{;}{a}{}{~}
\usepackage[]{geometry}
\usepackage{hyperref}
\begin{document}
\begin{center}
\Large{\textbf{Reality and Myths of AGN Feedback}}
\end{center}
\begin{center}
\large{Bernd Husemann${}^1$ \& Chris Harrison${}^2$}
\end{center}
${}^1$Max-Planck-Institut f\"ur Astronomie, K\"onigstuhl 17, D-69117 Heidelberg, Germany\\
\hspace*{3mm} \textit{email}:\href{mailto:husemann@mpia.de}{husemann@mpia.de}\\
${}^2$European Southern Observatory, Karl-Schwarzschild-Str. 2, D-85748 Garching b. M\"unchen, Germany\\

\textbf{Feedback from active galactic nuclei (AGN) remains controversial despite its wide acceptance as necessary to regulate massive galaxy growth. A dedicated workshop was held on 16-20 October 2017 at the Lorentz Center in Leiden to distinguish between the reality and myths of AGN feedback from the observational side. Here, we summarize briefly all the sessions and outcome of the stimulating workshop. More details on the outcome of the discussions are provided in a series of articles.}

Over the last few decades it has become the consensus that black holes, whose existence has recently been confirmed through the detection of gravitational waves \citep{Abbott:2016}, exist at the heart of every massive galaxy \citep{Kormendy:2013}. The accretion of matter onto these black holes, often weighing billions of solar masses, releases enormous amounts of energy across the electromagnetic spectrum. Twenty years ago, Silk \& Rees argued that, in principle, this energy could launch outflows of gas with sufficient velocities to become entirely unbound from their host galaxies. Such ``AGN feedback’’ can potentially significantly impact the evolution of the host galaxies. The seminal paper \citep{Silk:1998} is arguably responsible for an explosion of papers investigating the connection between AGN-driven outflows and galaxy formation \citep{Harrison:2018}. Since then, several observational results have emerged related to: the prevalence of AGN-driven outflows, their energetics, their driving mechanisms and their impact on star formation. However, contradictory conclusions have been presented in the literature on each of these topics. This motivated us to organize a workshop to understand these discrepancies.

\begin{figure}[ht!]
\includegraphics[width=\textwidth]{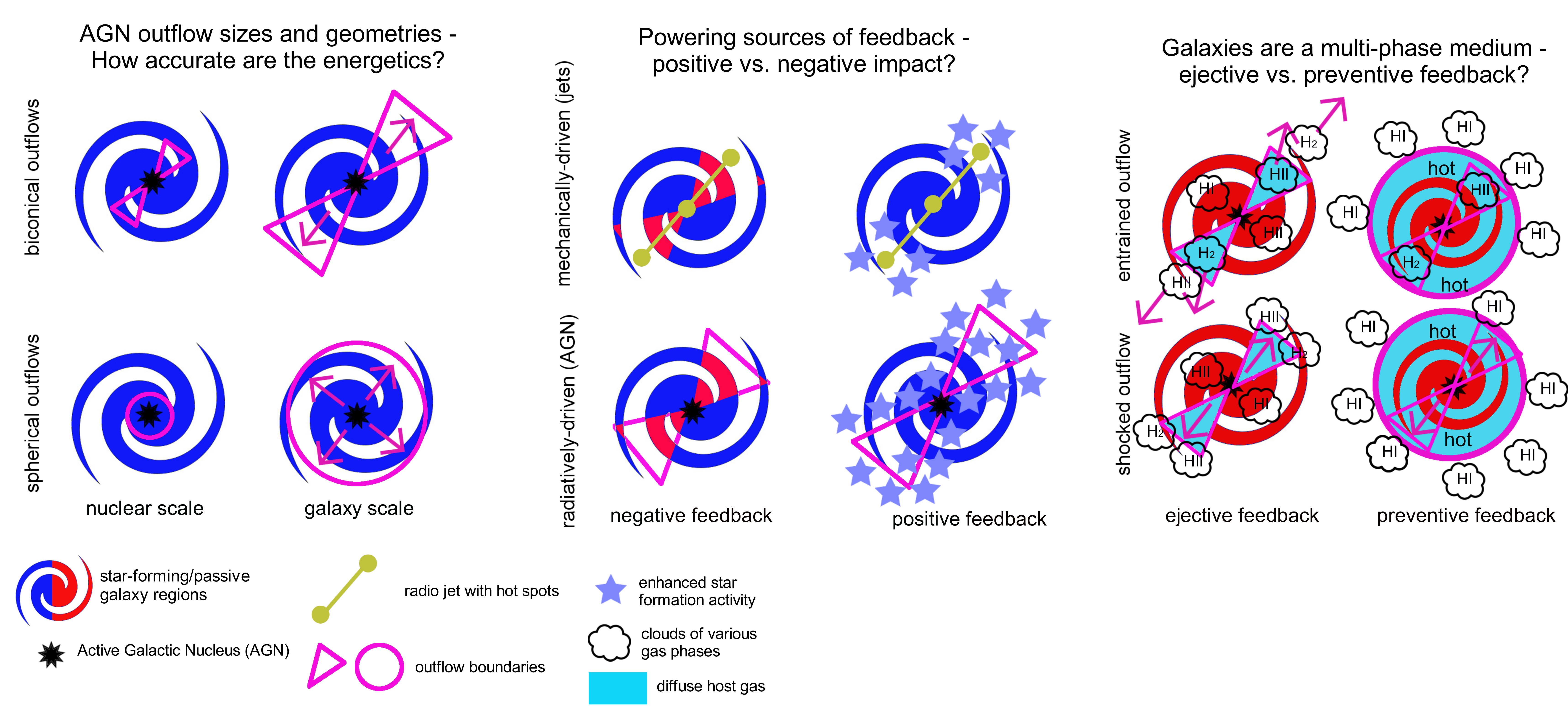}
 \caption{\small Cartoon representation of the import discussion points during the workshop. \textit{Left panel:} We discussed if current observations can differentiate between spherical and bi-conical outflows and if outflows are typically extended across the whole galaxy or are just located in the central regions. \textit{Middle panel:} We discussed if AGN outflows are radiatively-driven or mechanically-driven by a jet and if the outflows provide negative (reduction of star formation) or positive feedback (enhancement of star formation) to the host galaxies.  \textit{Right panel:} Multi-phase nature of the gas in galaxies play a key role to understand feedback process. We discussed if cold molecular gas (H${}_2$) and/or warm-ionized gas (HII)  is swept out of the galaxies in a confined shock-front or is entrained in the hot plasma outflow. A different aspect of this is whether negative AGN feedback is ejective, where gas is rapidly pushed out of the galaxy, or is preventive, where halo gas is kept hot and prevents cooling and condensation of atomic gas (HI) back to the galaxy and thereby stall formation of stars on longer timescales.}
 \label{fig:cartoon}
\end{figure}

27 participants convened at the Lorentz Center in Leiden between the 16 and 20 of October 2017. Every participant actively participated by contributing review or short science talks or by chairing break-out sessions. The interactive nature of the Lorentz Center enabled focused discussions and practical sessions. Below, we briefly summarize the discussions and main conclusions from our sessions addressing the key questions illustrated in Fig.~\ref{fig:cartoon}. Various Comments and a Perspective article with more in-depth discussions of the workshop outcomes are published in the same Nature Astronomy focus issue.

On the first day, Angela Bongiorno reviewed current observational campaigns to map large-scale ionized gas outflows. Bernd Husemann reviewed challenges in accurately measuring ionized gas outflow sizes and velocity structures due to the limited spatial resolution of observations. A related working session was dedicated to comparing different approaches to map the velocity structure and sizes of ionized gas outflows for published data sets (chairs: Bernd Husemann and Marios Karouzos).  We clearly established that spatially-resolved data contains more information than is often recovered using standard techniques. However, when applying deblending or forward modelling approaches, separate morpho-kinematic components can be recovered and carefully mapped. A parallel working session discussed various methods to constrain the electron densities in ionised outflows (chairs: Clive Tadhunter and Darshan Kakkad) - another key factor in deriving the physical properties of outflows. It was emphasized that assumed electron densities, in the absence of actual measurements, may often be two to three orders of magnitude lower than in reality. On the other hand, we currently lack good diagnostics for characterizing very low-density outflows which may also contain significant mass .

On the second day, Alessandro Marconi reviewed various assumptions, such as outflow geometry and plasma properties, that are made to derive energetics and mass outflow rates \citep{Harrison:2018}. In the related working session (chairs: Alessandro Marconi and Jong-Hak Woo) we concluded that outflow geometries in well-resolved cases in the nearby Universe appear to be mostly (hollow) bicones, in contrast to the sometimes assumed, expanding spherical shells. An identified action item was to degrade high-quality outflow observations in spatial resolution and signal-to-noise to match representative observations of high-redshift galaxies. This could provide preliminary calibrations for recovering key physical parameters in lower quality data. The same day, Claudia Cicone reviewed the observations on the multi-phase nature of outflows. The crucial message is that all phases contribute to the mass and energy content and we currently have limited knowledge on the relative contributions for unbiased galaxy populations. The related working session (chairs: Claudia Cicone and Marcella Brusa) discussed the very few observations where multi-phase AGN-driven outflows have been observed. It became clear that systematic multi-frequency observations of unbiased and large samples using ALMA, SINFONI, MUSE and other new facilities are necessary to develop a comprehensive physical picture of AGN outflows. However, these unbiased searches for multiphase outflows would require a significant investment of observational resources and time from the community \citep{Cicone:2018}.

The third day focused on the impact of AGN on their host galaxies. Raffaella Morganti reviewed the observations available of outflows driven by radio jets and the impact these can have on the host galaxy. Giovanni Cresci reviewed the evidence for positive and negative feedback driven by outflows in some well-studied quasars. The participants explored the related issues in two working sessions on: (1) the powering mechanisms of outflows and (2) the diagnostics for positive versus negative AGN feedback. The highlight of the former session (chairs: Dominika Wylezalek and Raffaella Morganti) was a quick proposal design and an on-the-fly evaluation by participants. A consensus was reached that deep and high spatial resolution radio observations are required to systematically explore the drivers of outflows \citep{Wylezalek:2018}. The second session (chairs: Roberto Maiolino and Giovanni Cresci) discussed various feedback scenarios. They concluded that a delayed “preventive” mode of feedback is more likely to be the dominant mechanism for the suppression of star formation compared to a fast “ejective” mode, where star forming material is rapidly removed from the host galaxy \citep{Cresci:2018}.

The discussion on the impact of AGN feedback continued the next day. Chris Harrison reviewed observational results on the impact of AGN on star formation in the overall galaxy population. In the corresponding working session (chairs: Chris Harrison and Angela Bongiorno), small groups were given several papers each to understand apparently discrepant conclusions. A consensus was reached that there is currently no strong direct observational evidence for the impact of AGN on star formation in the overall galaxy population when different approaches and selection effects are taken into account \citep{Harrison:2017}. The same day Bernd Husemann highlighted the capabilities of the upcoming James Webb Space Telescope and described the current plans of guaranteed time observations related to AGN feedback. 

Finally, Jong-Hak Woo summarized the discussions and conclusions made during the week. Exploiting new facilities for systematic multi-wavelength studies of unbiased galaxies samples and working more closely with theorists were identified as key for future progress. Thanks to the efficient organization by Lorentz Centre staff and the active contribution of all participants, the workshop was a great success, enlightening PhD students, postdoctoral researchers and senior astronomers alike. Despite the great challenges ahead, we left with clear ideas for future promising avenues to explore as described in the related articles in the corresponding Nature Astronomy focus issue.\bigskip\\

\textbf{Acknowledgments:}
The great success of this workshop was only possible due to the enormous effort of the scientific organizing committee, Bernd Husemann, Angela Bongiorno, Chris Harrison, Vincenzo Mainieri and Raffaella Morganti as well as the engagement of all participants. All of the review talks mentioned in this report are available at the \href{http://www.lorentzcenter.nl/lc/web/2017/912/info.php3?wsid=912}{Lorentz-Center web page of this workshop}. Special thanks to Nienke Tander for the fantastic logistical organization of the meeting and conference dinner. We also acknowledge the substantial financial support by the Lorentz Center for the workshop.

\bibliography{references}

\end{document}